\documentclass[aps,prx,twocolumn,preprintnumbers,amsmath,amssymb,superscriptaddress]{revtex4-1}
\usepackage[english]{babel}
\usepackage{graphicx}
\usepackage{stmaryrd}
\usepackage{amssymb}
\usepackage{amsfonts}
\usepackage{amsmath}
\usepackage{dsfont}
\usepackage{color}
\usepackage{xspace}
\usepackage{multirow}
\usepackage{comment}
\usepackage{url}
\usepackage{hyperref}

\usepackage{bm}
\usepackage{amsthm}

\DeclareSymbolFont{largesymbols}{OMX}{cmex}{m}{n}

\def\be{\begin{equation}}       \def\ee{\end{equation}}
\def\bea{\begin{eqnarray}}      \def\eea{\end{eqnarray}}
\def\ba{\begin{array}}
\def\ea{\end{array}}

\DeclareGraphicsExtensions{.pdf,.ps,.eps}

\begin{document}

\title{Flat-band Ferromagnetism of SU$(N)$ Hubbard Model on Tasaki Lattices}

\author{Ruijin Liu}
\affiliation{Department of Physics, Renmin University of China, Beijing 100872, China}

\author{Wenxing Nie}
\email{wenxing.nie@gmail.com}
\affiliation{Center for Theoretical Physics, College of Physics, Sichuan University, Chengdu, Sichuan 610064, China}

\author{Wei Zhang}
\email{wzhangl@ruc.edu.cn}
\affiliation{Department of Physics, Renmin University of China, Beijing 100872, China}
\affiliation{Beijing Key Laboratory of Opto-electronic Functional Materials and Micro-nano Devices, Renmin University of China,
Beijing 100872, China}

\date{\today}

\begin{abstract}
We investigate the para-ferro magnetic transition of the repulsive SU($N$) Hubbard model on a type of one- and two-dimensional decorated cubic lattices, referred as Tasaki lattices, which feature massive single-particle ground state degeneracy. Under certain restrictions for constructing localized many-particle ground states of flat-band ferromagnetism, the quantum model of strongly correlated electrons is mapped to a classical statistical geometric site-percolation problem, where the nontrivial weights of different configurations must be considered. We prove rigorously the existence of para-ferro transition for the SU($N$) Hubbard model on one-dimensional Tasaki lattice and determine the critical density by the transfer-matrix method. In two dimensions, we numerically investigate the phase transition of SU($3$), SU($4$) and SU($10$) Hubbard models by Metropolis Monte Carlo simulation. We find that the critical density exceeds that of standard percolation, and increases with spin degrees of freedom, implying that the effective repulsive interaction becomes stronger for larger $N$. We further rigorously prove the existence of flat-band ferromagnetism of the SU$(N)$ Hubbard model when the number of particles equals to the degeneracy of the lowest band in the single-particle energy spectrum.
\end{abstract}

\maketitle

\textbf{1.Introduction}\\

In condensed matter physics, the interplay of the Coulomb interaction with Pauli principle is a fundamental problem, from where many intriguing phases can appear. One important problem is the emergence of ferromagnetism in two or higher dimensions, which can date back to the early age of quantum theory and reveal the importance of many-body effect. After more than one century of investigation, a few special cases have been solved by either exact results~\cite{Nagaoka,Lieb-theorem,Mielke1,Mielke2,TasakiPRL,Mielke-Tasaki,Tasaki-review,YiLi,Li-Nagaoka} or sign-problem-free quantum Monte Carlo simulation~\cite{Li-QMC-FM}. However, a general solution to this problem is still lack, partly due to the notorious sign problem of fermions~\cite{book-Lieb,book-Flliott}. As an interesting and important limiting case, a system with a flat band in the single-electron energy spectrum may favor the stabilization of ferromagnetism as it reduces the kinetic energy cost for spin polarization to zero due to the band flatness. For instance, Mielke and Tasaki prove flat-band ferromagnetism in a class of lattices, based on a systematic route called ``cell construction''~\cite{Mielke1,Mielke2,TasakiPRL,Mielke-Tasaki}. After that, flat-band ferromagnetism has attracted a lot of attention and has been connected with the discussion of topological and magnetic phases~\cite{ShengNatComm,SunPRL, Tang2011,Katsura-Tasaki,Scaffidi,Essafi,Tsai,Goldman,Shizhong-percolation}.

The previous investigation on flat-band ferromagnetism mainly focus on fermionic systems possessing the SU($2$) spin-rotational symmetry. The SU($N>2$) generalization of flat-band ferromagnetism is not only of theoretical interests~\cite{Affleck1,Affleck2,Sachdev1,Sachdev2,Wu-SO5,Wu-NatPhy,Wu-view,Cazalilla,NiePRA}, but also of experimental relevance in the context of cold atomic gases ~\cite{Takahashi_PRL,Takahashi_Nat,Taie,Bloch}, where the high degrees of freedom of nuclear spin provides a fascinating playground to simulate the SU($N$) fermionic Hubbard model~\cite{Cappellini,ZhangSci,Pagano2014,Scazza}. By loading ultracold alkaline-earth-metal(-like) atoms in different optical lattices, the quantum simulation of various models has been theoretically proposed and further experimentally realized~\cite{Sr-Rice,Fallani,Takahashi_PRL,Takahashi_Nat}, where the spin symmetries can be as high as SU($6$) for ${}^{173}$Yb~\cite{TaiePRL20010,Takahashi_Nat,Taie,Hofrichter} and SU($10$) for ${}^{87}$Sr~\cite{Sr-Rice,Sr-Desalvo}.
\begin{figure}[t]
	\begin{center}
		\includegraphics[width=5cm]{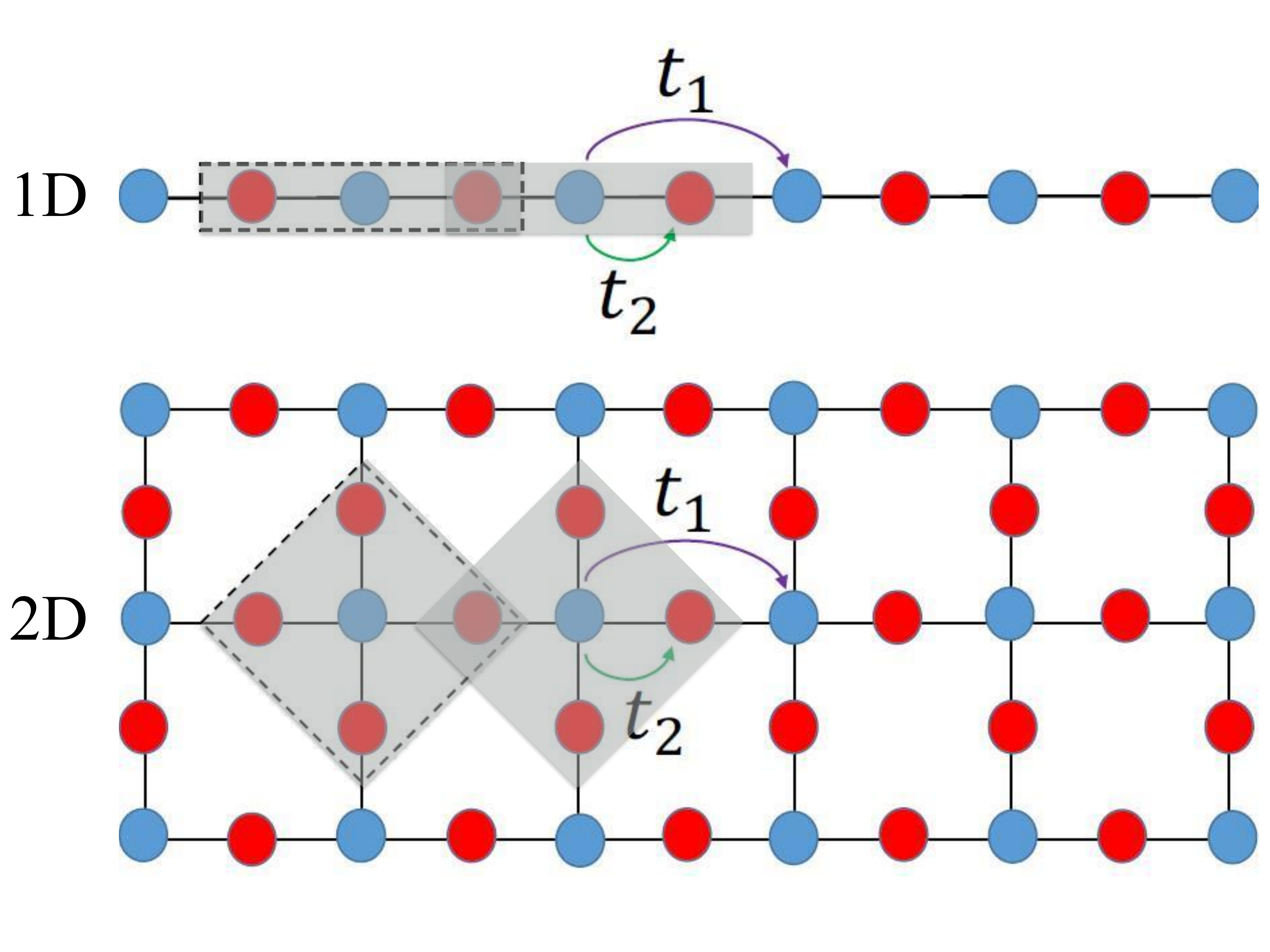}
	\end{center}
	\caption{\label{lattice}(Color online) An illustration of the decorated cubic lattices designed by Tasaki in $1$D and $2$D: blue sites denote the undecorated hypercubic lattice, and the red ones denote the decorated sites. A trapping cell indicated by dashed-line area contains three (five) sites for $1$D ($2$D) Tasaki lattice. The shaded area in grey shows an example of cluster of length two, with the overlapping of two adjacent trapping cells.}
\end{figure}

In this paper, we investigate the para-ferro magnetic transition and flat-band ferromagnetism of the SU($N$) Hubbard model on a special class of decorated cubic lattices in one and two dimensions, which are illustrated in Fig.~\ref{lattice}. The existence of ferromagnetism with SU($2$) spin symmetry on this lattice, later referred as Tasaki lattice for short, is first considered by Mielke and Tasaki~\cite{Mielke1,Mielke2,TasakiPRL,Mielke-Tasaki}, and further by Maksymenko {\it et al.}~\cite{MaksymenkoPRL}, who introduced a mapping from the quantum Hubbard model to a classical statistical geometric site-percolation problem. Here, we first prove a generalized version of this mapping from the SU($N$) Hubbard model to an $N$-state Pauli-correlated percolation (PCP) problem, where an SU($N$) correlated weight for each configuration is considered. Then we obtain the exact results in one-dimensional ($1$D) Tasaki lattice for the SU($N$) Hubbard model, and further verify the para-ferro magnetic transition numerically. For the two-dimensional ($2$D) case, we implement importance samplings by Metropolis Monte Carlo simulation and determine the para-ferro magnetic transition numerically. Moreover, we prove rigorously the existence of flat-band ferromagnetism of the SU$(N)$ Hubbard model, when the number of particles $n$ equals to the degeneracy $\mathcal{N}$ of the lowest band in the single-particle energy spectrum.\\

\textbf{2.Mapping from the SU($N$) Hubbard model to an $N$-state Pauli-correlated percolation problem}\\

We start with the  SU$(N)$ Hubbard model with repulsive interaction $U>0$ on a Tasaki lattice $\Lambda$,
\be
H=\sum_{i,j \in \Lambda}t_{i,j} \left( c_{i,\sigma}^{\dag}c_{j,\sigma}+\mbox{H.c.} \right) + U\sum_{i} \sum_{\sigma\neq\sigma'} n_{i,\sigma}n_{i,\sigma'},\\
\label{hamiltonian}
\ee
where ${\sigma} = \{1,\cdots,N \}$ labels the spin (or equivalently flavor or color) of a fermion. The operator $c_{i,\sigma}$ and its adjoint $c_{i,\sigma}^{\dag}$ satisfy the fermionic anticommutation relations $\{ c_{i,\sigma}^{\dag}, c_{j,\sigma'}\}=\delta_{ij}\delta_{\sigma\sigma'}$ and $\{ c_{i,\sigma}, c_{j,\sigma}\}=\{ c_{i,\sigma}^{\dagger}, c_{j,\sigma}^{\dagger}\}=0$, and the number operator $n_{i,\sigma} \equiv c_{i,\sigma}^{\dagger} c_{i,\sigma}$. The Tasaki lattice $\Lambda$ is a $d$-dimensional hypercubic lattice with decorated sites located at the middle of the nearest links of the hypercubic lattice~\cite{Tasaki-review}. The hopping pattern, therefore, is composed by the nearest-neighbor ones ($t_1$) on the original hypercubic lattice and the nearest-neighbor ones  ($t_2$) after the decoration, as shown in Fig.~\ref{lattice}.

The lowest band of single-particle energy spectrum of the Tasaki lattice is completely dispersionless as long as  $t_2=\sqrt{c}\, t_1>0$, in which $c$ is the coordination number of the corresponding undecorated lattice ($c=2$ and $c=4$ for $1$D and $2$D lattices, respectively).
The existence of flat band is due to destructive interferences of particle hoppings, implying that we can construct the localized single-particle eigenstate in each trapping cell~\cite{Schulenburg-PRL2002,
Zhitomirsky-Tsunetsugu,Derzhko2007summary,DerzhkoPRB2007}, whose wave function only overlaps with that of particles in adjacent cells, as shown by dark grey areas in Fig.~\ref{lattice}. When non-interacting particles are filled into the flat band, the ground states are highly degenerate respective to the spin of individual particles. The ferromagnetic state with all particles possess the same spin is obviously one of these degenerate ground states.

\begin{figure}[t]
	\begin{center}
		\includegraphics[width=7cm]{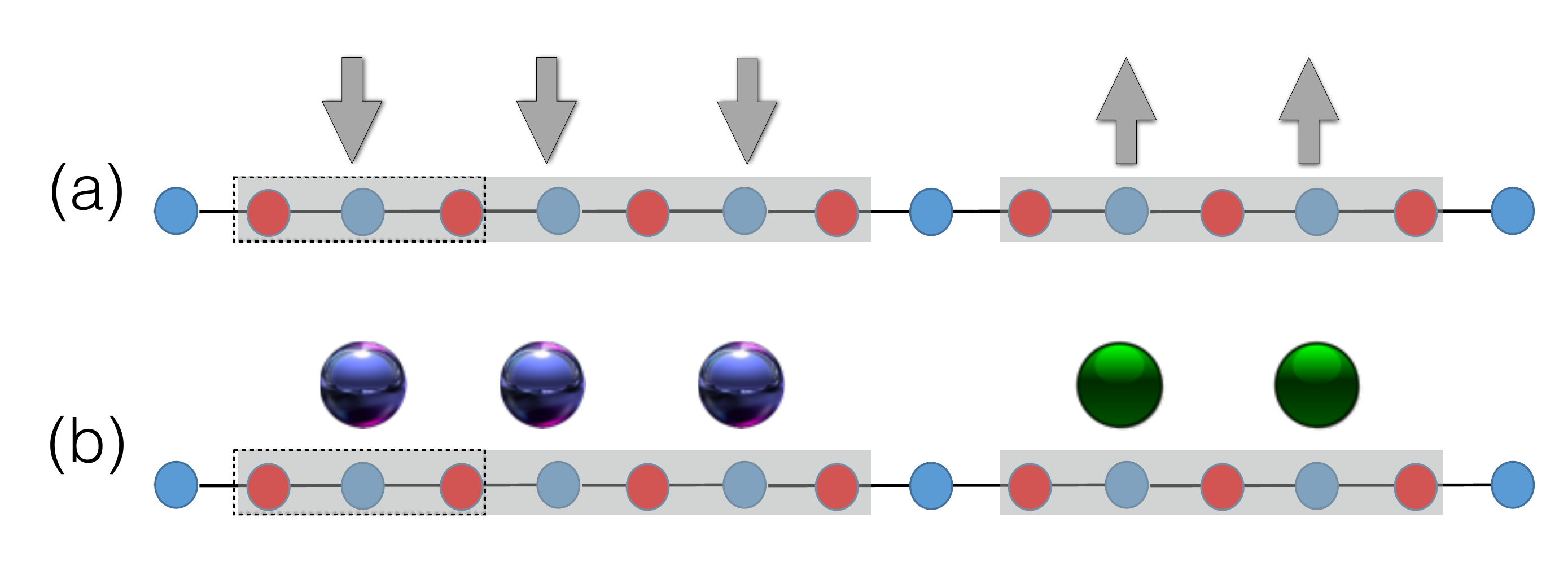}
	\end{center}
	\caption{\label{mapping}(Color online) An illustration of the mapping between (a) the ferromagnetic states and (b) the configuration of linked clusters of percolation of $2$-color particles. The shaded areas in grey show examples of ferromagnetic clusters in (a) and linked occupied trapping cells in the perspective of percolation in (b). A trapping cell is illustrated by the area enclosed by the dashed-lines.}
\end{figure}

Since the repulsive interaction is positive semidefinite, the ground state energy may not decrease as one turns on $U>0$. In the limit of low particle density, each trapping cell is occupied by at most one particle to avoid energy penalty. When the density increases, the wave function of a trapping cell can not avoid overlapping with that of adjacent one. But the energy penalty can be avoided as well if the particles in neighboring trapping cells are in the same symmetric spin state~\cite{MaksymenkoPRL}. By doing so, particles on these linked trapping cells are aligned in spin, and form a polarized ``cluster'' as shown in Fig.~\ref{mapping}a. Namely, the minimization of energy requires all spins of particles belong to the same cluster being aligned, and this is the origin of ferromagnetism.

We then establish a one-to-one correspondence between the states composed of ferromagnetic clusters and the geometric configurations of $n$ $(\leq\mathcal{N})$ particles distributed over $\mathcal{N}$ traps. In comparison to the standard percolation~\cite{Stauffer, Isichenko}, the ferromagnetic state of a cluster $C$ of size $|C|$ (or the number of particles in cluster $C$) acquires an additional spin degeneracy, which is given by the dimension of the irreducible representation of the SU$(N)$ group~\cite{Georgi-book,Katsura},
\begin{align}
d_{\text{SU}(N)}(|C|)=(N+|C|-1)!/(|C|!(N-1)!).
\label{degeneracy}
\end{align}

For example, the degeneracy of the ferromagnetic state in the SU($2$) Hubbard model is $2S+1$, where $S$ is the total spin of this cluster. Therefore, for a ferromagnetic cluster of length three and with total spin $S=3/2$ as shown on the left side Fig.~\ref{mapping}a, we need to assign a factor of $4$ to the geometric configuration of the cluster composed by three linked trapping cells depicted on the left of Fig.~\ref{mapping}b.  Note that the irreducible representation is the ``permutation symmetric'' representation of the SU$(N)$ group. As the ferromagnetic cluster carries the fully symmetric representation of the SU($N$) group, the electron wave function is fully antisymmetric in the spatial space, thereby avoiding the interaction energy penalty. Thus, for a given configuration $q$ composed by a set of linked clusters, a nontrivial weight $W(q)$ must be assigned by taking into account the spin degeneracy of each cluster
\begin{align}
W(q)=\prod_{i=1}^{M_q}e^{\mu |C_i|}d_{\text{SU}(N)}(|C_i|),
\label{weight}
\end{align}
where $e^\mu$ is the fugacity to tune the number of particles in a grand-canonical ensemble. Further analysis shows that for a given number of particles, the relative probability for them to decompose into smaller clusters is higher than forming larger clusters, which implies that this nontrivial weight gives rise to an effective repulsive interaction, tending to break up large clusters~\cite{MaksymenkoPRL}. In the following model calculation, it will be shown that the repulsive effect becomes stronger with $N$, such that the formation of clusters with size proportional to the system size, i.e., the percolation transition, becomes more difficult in systems with higher SU$(N)$ spin symmetry.

The key theorem then needs to be proved is that the ground states of the SU$(N)$ Hubbard model when the flat band is partially filled can be represented by the states of ferromagnetic clusters, which can further be mapped into geometric configurations (see Supplementary Material accompanying
this paper). Then the quantum many-body system of SU$(N)$ flat-band ferromagnetism is reduced to a classical site-percolation problem with a nontrivial weight. The expectation value of an operator $O$ is the average over all degenerate ground states,
\begin{align}
\langle O\rangle=\frac{\sum\limits_{q}O_qW(q)}{\sum\limits_{q}W(q)}.
\label{expectation}
\end{align}
For the sake of magnetism, we consider the total spin operator $\mathbf{S}$ and focus on the expectation value $\langle \mathbf{S}^2 \rangle = \sum_q \mathbf{S}_q^2 W(q)/\sum_q W(q)$, where $\mathbf{S}^2$ is the quadratic Casimir operator of the SU$(N)$ group. Because the particles belonging to the same cluster are aligned in the ferromagnetic state, the eigenvalue of $\mathbf{S}^2$ for a cluster with size $|C|$ is~\cite{Ma-book}
\begin{align}
S^2(|C|)=\frac{(N-1)(N+|C|)|C|}{2N}.
\label{sc}
\end{align}
The total spin $\mathbf{S}_q^2$ for a given geometric configuration $q$ can be calculated in two equivalent ways
\begin{align}
S_q^2&=\sum_{i=1}^{M_q}S^2(|C_i|)\nonumber\\
&=\sum_{l=1}^{n}N_q(l)S^2(l)=\mathcal{N} \sum_{l=1}^{n}n_q(l)S^2(l),
\label{sq}
\end{align}
where $N_q(l)$ is the number of clusters of size $l$, and $n_q(l) \equiv N_q(l)/\mathcal{N}$ is the normalized fraction with respect to the number of traps $\mathcal{N}$. The average value of the fraction $n(l) = \sum_q n_q(l) W(q) / \sum_q W(q)$ is a characteristic quantity in percolation as it signals the information of cluster size distribution~\cite{Stauffer, Isichenko}.\\

3.\textbf{Exact results in one dimension}\\

As we have discussed in the previous section, the many-body ground states of the SU$(N)$ Hubbard model on flat bands can be mapped to a geometric site-percolation problem with a nontrivial weight. In this section, we consider
the 1D Tasaki lattice as an example, and present some exact results obtained by the transfer-matrix method~\cite{MaksymenkoPRL}.
The most important step is to construct a one-to-one mapping between the enumeration of all ground states of the Hubbard model and the counting of configurations when populating $n$ particles with spin component $1,2,\cdots,N$ into $\mathcal N$ trapping cells. Here, we establish such a mapping by introducing the following rules to populate particles with spin component $\sigma=1,2,\cdots,N$ to the trapping cells~\cite{DerzhkoPRB2010}:

(1) Each trap $j$ can be either empty or occupied by only one particle with spin component $\sigma$. In other words, for each trap $j=0,1, \cdots, \mathcal{N}-1$, there are $N+1$ trap states $\sigma_j=0,1,\cdots,N$, where $0$ indicates the trap is empty. Then a ground state of the chain is presented as a sequence of the trap states.

(2) We take the convention $\sigma_i \le \sigma_{i+1}$ to avoid over counting, because only one sequence of two trap states corresponding to the neighboring cells being occupied by different spin-component particles is allowed when moving along adjacent traps.

After employing the two
rules above, we can construct a transfer matrix $\mathbf{T}$ to represent the percolation problem, where the grand partition function is given by $\Xi(z,\mathcal{N})= \text{Tr} \, \mathbf{T}^{\mathcal N}$ (see Supplementary Material accompanying
this paper).
Figure~\ref{1ds2cluster} shows the cluster distribution $n(l)$ and the macroscopic magnetic moment $S^2/S^2_{\text{max}}$ of the Hubbard model with SU($3$), SU($4$) and SU($10$) symmetries for instances, at the particle density $p \equiv n / \mathcal{N} =0.99$.
For the cluster distribution $n(l)$ depicted in Fig.~\ref{1ds2cluster}a, a non-monotonic behavior with a peak structure at some finite $l$ and vanishing value for large enough cluster is observed for all cases. With increasing $N$, the peak becomes higher and narrower, with the central position shifted to a larger value of $l$. This tendency indicates that the effective repulsion induced by the nontrivial weight is more prominent for cases of higher spin symmetry. From Fig.~\ref{1ds2cluster}b, we find that the macroscopic magnetic moment scales to zero for all cases, implying that the ground state is still paramagnetic when the flat band is nearly $1/N$ filled. Besides, the macroscopic magnetic moment for SU$(10)$ is smaller than that of SU$(3)$ at the same concentration. This observation suggests that the formation of large enough clusters is more difficult in SU$(10)$ system than the SU$(3)$ case, as can also be inferred from the tails on the right side of Fig.~\ref{1ds2cluster}a.
We will find these results consistent with following analytic discussions.

\begin{figure}
	\begin{center}
		\includegraphics[width=7cm]{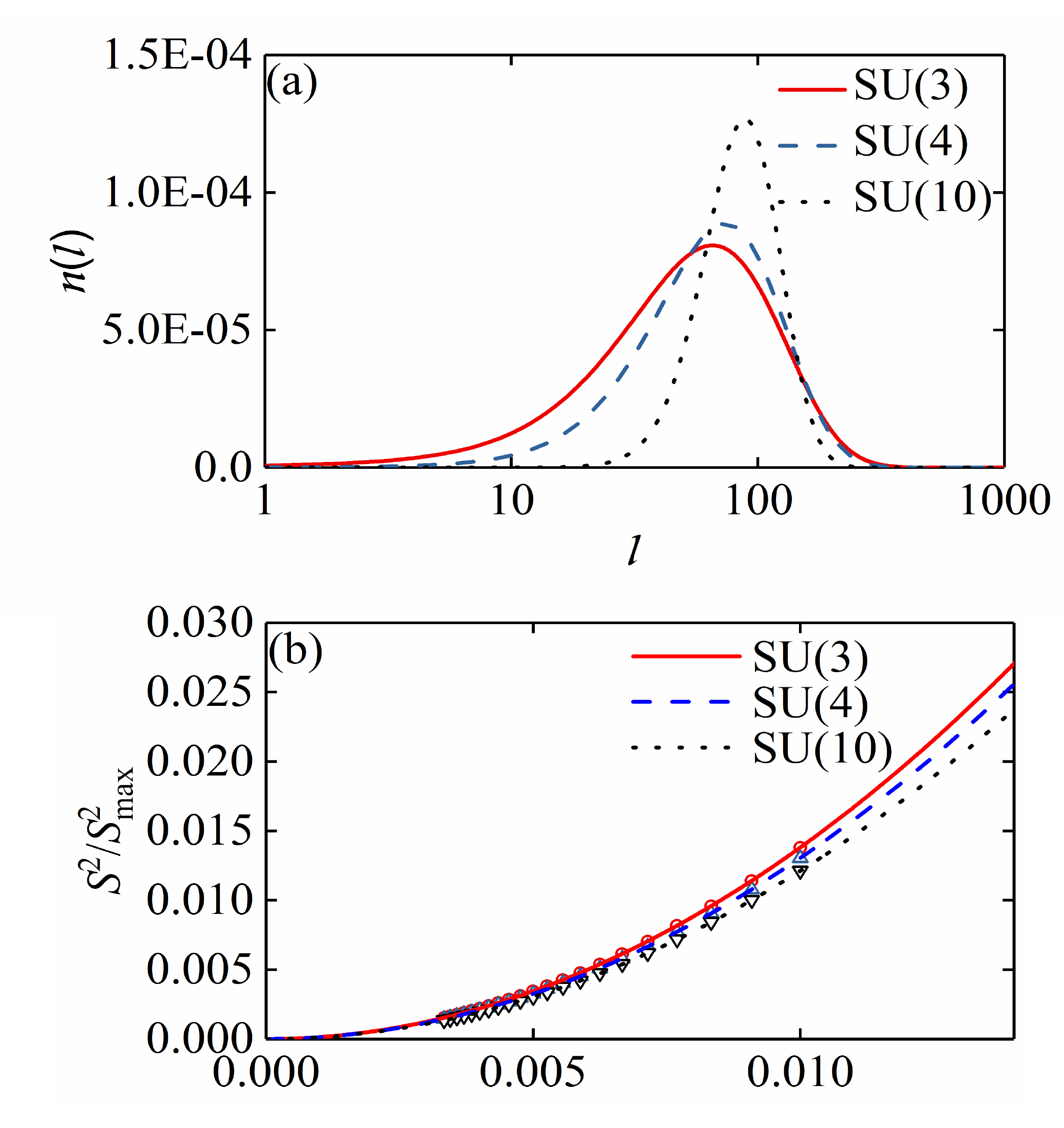}
	\end{center}
	\caption{\label{1ds2cluster}(Color online) (a) Cluster distributions and (b) macroscopic magnetic moments in $1$D Tasaki lattice for the Hubbard model with SU$(3)$, SU$(4)$ and SU$(10)$ symmetries. The particle density $p=0.99$ and the number of trapping cells $\mathcal{N}=10^5$ are chosen for calculation in all cases. The macroscopic magnetic moments $S^2_{\textrm{max}}$ is determined by arranging all particles within one single cluster.}
\end{figure}
\begin{figure*}
	\begin{center}
		\includegraphics[width=6.5in]{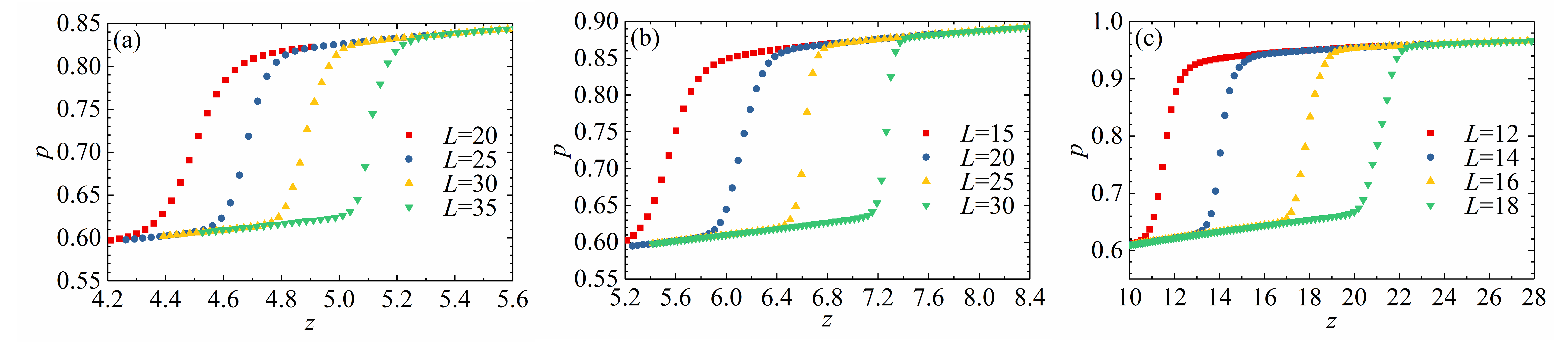}
	\end{center}
	\caption{\label{2dpz}(Color online) Particle density $p$ versus fugacity $z$
	in $2$D Tasaki lattice for the Hubbard model with (a) SU$(3)$, (b) SU$(4)$ and (c) SU$(10)$ symmetries. }
\end{figure*}
\begin{figure*}
\centering
\includegraphics[width=6.5in]{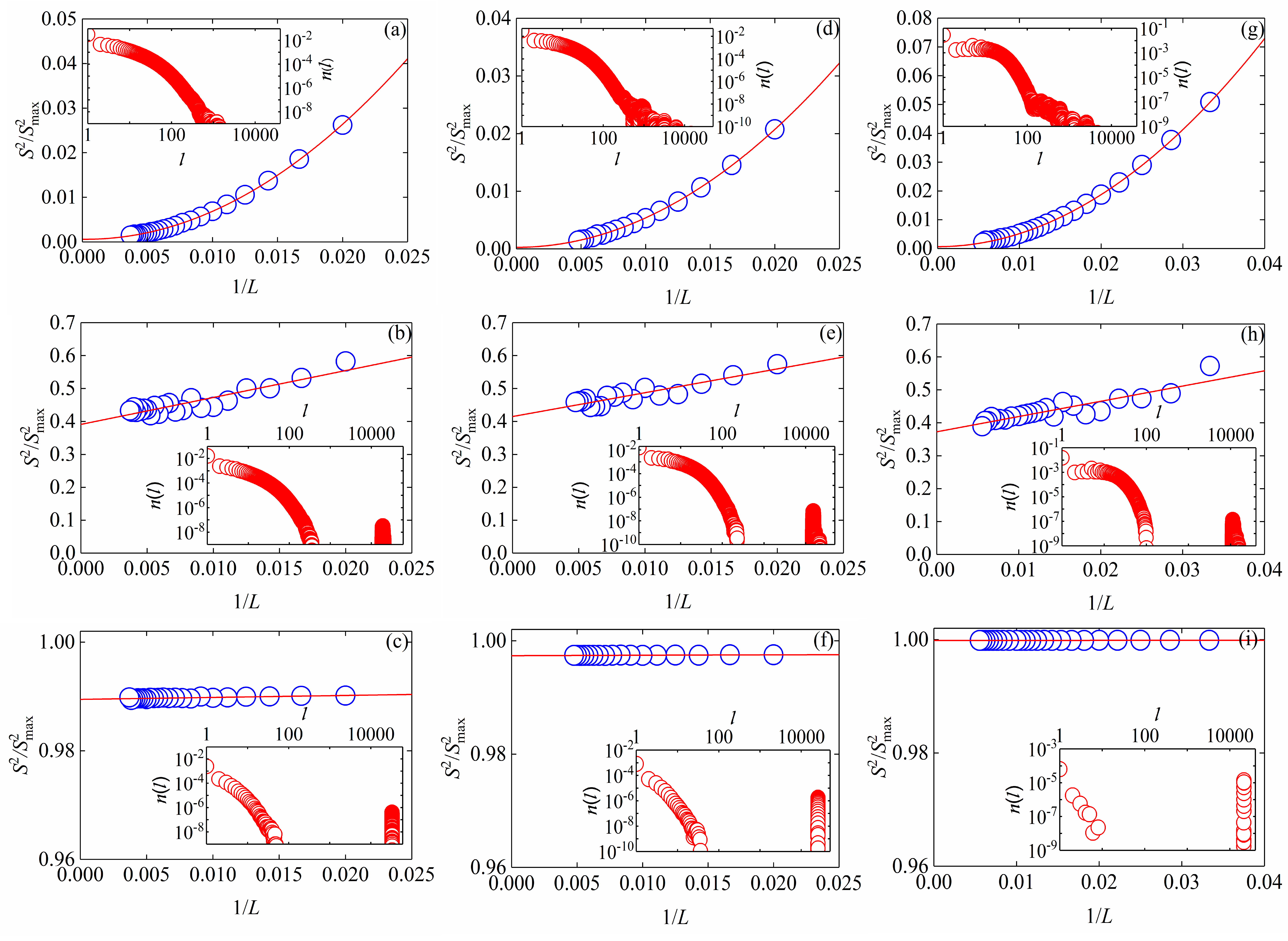}
\caption{(Color online) Macroscopic magnetic moment $S^2/S_{\text{max}}^2$ for 2D Tasaki lattice. For the SU$(3)$ case, numerical results (green circles) are depicted for system size up to $L=270$ at particle densities (a) $p=0.64$, (b) $p=0.74$, (c) $p=0.83$. For the SU$(4)$ case, lattice size is up to $L=210$ with particle densities (d) $p=0.65$, (e) $p=0.77$, (f) $p=0.88$. For the SU$(10)$ case, lattice size is up to $L=200$ and the particle densities are chosen as (g) $p=0.69$, (h) $p=0.82$, (i) $p=0.95$. The maximal magnetic moment $S^2_{\text{max}}$ is calculated by arranging all particles in one single cluster. The insets show cluster distributions with $L=200$ for SU$(3)$, $L=170$ for SU$(4)$, and $L=150$ for SU$(10)$, respectively.}
\label{2dmagnetic}
\end{figure*}

In the thermodynamic limit ($\mathcal{N}\to\infty$ with $n/\mathcal{N}$ finite), we obtain the cluster distribution
\begin{align}
n(l,\text{SU}(N))= f_{\text{SU}(N)}(z)\frac{(l+N-1)!}{l!} \left(\frac{z}{\lambda_{\text{max}}(z)} \right)^l, \label{cluster}
\end{align}
where $z=e^\mu$ is the fugacity, $f_{\text{SU}(N)}(z)$ is a function depends only on $z$ and $N$, and $\lambda_{\text{max}}(z)$ is the eigenvalue with maximal modulus of the $\mathbf{T}$-matrix. The exact expression of $f_{\text{SU}(N)}(z)$ is in general complicated, but fortunately we can treat it as a coefficient function such that some important properties of the system, including the shape of the cluster distribution $n(l)$, can be obtained qualitatively. From Eq.~(\ref{cluster}), one can easily conclude that as long as ${z/\lambda_{\text{max}}(z)}<1$, clusters of large enough size must vanish. The percolation transition thus occurs when the condition ${z_c/\lambda_{\text{max}}(z_c)}=1$ is satisfied, which results in $z_c \to +\infty$ (see Supplementary Material accompanying
this paper).

For the macroscopic magnetic moment, we have
\begin{align}
\frac{S^2}{\mathcal{N}^2}=\sum_{l=1}^{n}\frac{S_{l}^{2}}{\mathcal{N}}f_{\text{SU}(N)}(z)\frac{(l+N-1)!}{l!} \left(\frac{z}{\lambda_{\text{max}}}\right)^l,
\label{magmoment}
\end{align}
which also proves the ground state is paramagnetic for finite $z$. Further calculation shows that in the thermodynamic limit, the particle density $p$ indeed approaches unity as $z \to +\infty$. Hence the ground state is paramagnetic at any particle density $p<1$ for arbitrary $N$, with an onset of ferromagnetism at $p=1$. These results are consistent with our rigorous proof of flat band ferromagnetism for SU$(N)$ Hubbard model on the Tasaki lattice at filling $p=1$ (see Supplementary Material accompanying
this paper). One should remind that the conclusions stated above does not contradict with the Lieb-Mattis theorem~\cite{Lieb-Mattis}, which proves the absence of ferromagnetism of 1D Hubbard model on lattices without loops. However, here we consider a special type of lattices with loops composed by nearest $t_1$ and next-nearest neighbor hoppings $t_2$ as depicted in Fig.~\ref{lattice}. In fact, as the existence of loop is crucial to validate the construction of the many-body ground states~\cite{Tasaki-review,Nie2013,Nie2018}, the Lieb-Mattis theorem does not apply in this circumstance.\\

4.\textbf{Para-ferro magnetic transition in two dimensions}\\

For 2D Tasaki lattice, the transfer matrix formalism employed above can not be directly applied~\cite{2DTMM}. For instead, we make use of the Metropolis Monte Carlo method with importance samplings for the nontrivial weights to simulate the percolation problem on $2$D $L\times L$ Tasaki lattices with periodic boundary conditions. A new configuration is accepted with the Metropolis probability min$[1,W(q')/W(q)]$, in which $q$ denotes the old configuration and $q'$ denotes a new configuration. Specifically, to obtain the density of particles versus fugacity $z=e^{\mu}$, we perform grand-canonical simulations with exchange Monte Carlo. We randomly choose a site and change its occupying status, i.e., if it is (not) occupied, we (populate) remove one particle.

The percolation transition can be identified as a first order jump between concentrations $p_{-}$ and $p_{+}$ shown in Fig.~\ref{2dpz}. For fillings below $p_{-}$, the possibility to find a cluster of the same color is vanishingly small in the thermodynamic limit, suggesting a paramagnetic state in the original Hubbard model. When $p > p_{+}$, the clusters of the same color can be linked together to reach a macroscopic size. In the SU($N$) Hubbard model, this regime corresponds to a system with a macroscopic ferromagnetic domain.

In order to determine the para-ferro magnetic transition point, we perform canonical simulations with fixed density of particles for various lattice sizes $L$ to obtain the macroscopic magnetic moments, and extrapolate the results to the thermodynamic limit. Numerically, we generate a new configuration $q'$ from the old one $q$ by randomly permuting the occupied status of two sites to guarantee the constraint that the number of particles is fixed. As shown in Figs.~\ref{2dmagnetic}a, \ref{2dmagnetic}d and \ref{2dmagnetic}g, the macroscopic magnetic moments scale to zero at $p=0.64$ for the SU$(3)$ Hubbard model, at $p=0.65$ for the SU$(4)$ model, and at $p=0.69$ for the SU$(10)$ case. This implies that the corresponding ground states are paramagnetic below such fillings. In Figs.~\ref{2dmagnetic}b, \ref{2dmagnetic}e and \ref{2dmagnetic}h, the particle densities are increased to $p=0.74$, $0.77$, and $0.82$ for the SU$(3)$,  SU$(4)$, and SU$(10)$ models, respectively. The macroscopic magnetic moments in all cases scale to a finite values in the thermodynamic limit, indicating the emergence of ferromagnetism. In this regime, the snapshot (see Supplementary Material accompanying
this paper) shows the concomitant phase-separated states, where macroscopic ferromagnetic domains can be observed. In Figs.~\ref{2dmagnetic}c, \ref{2dmagnetic}f and \ref{2dmagnetic}i, the particle density is further enhanced to $p=0.83$, $0.88$, and $0.95$ for the SU$(3)$,  SU$(4)$, and SU$(10)$ cases, respectively. The macroscopic magnetic moments scale to finite values with $S^2/S_{\text{max}}^2 \approx 1$ in the thermodynamic limit, and are independent of the system size. In this regime, the snapshot shows the ferromagnetic phase spans across the entire system.

Based on the numerical results in both grand canonical and canonical ensembles, we obtain the phase diagrams for SU$(N)$ Hubbard models on 2D Tasaki lattice. By increasing particle density $p$, the para-ferro transition as a first order jump is found to lay between concentrations $p_{-}$ and $p_{+}$, whose exact values are both enhanced with $N$. When $p<p_{-}$, the ground state is paramagnetic with zero macroscopic magnetic moment. When $p_{-}<p<p_{+}$, the ground state falls into the phase-separation regime of ferromagnetic domains. As $p_{+}<p<1$, the ground state is of unsaturated ferromagnetism, where the ferromagnetic phase spans across the entire system. The dependence of the exact values of $p_{-}$ and $p_{+}$ on the value of $N$ is rooted from the stronger effective repulsion in systems with larger $N$. Finally, as the flat band is $1/N$ filled with $p=1$, the ground state for arbitrary $N$ is rigorously proved to be the ferromagnetic state (see Supplementary Material accompanying
this paper).

It is interesting to consider the limit $N\to\infty$, where a numerical calculation becomes difficult by directly using the method discussed above due to the largely extended space of spin degrees of freedom. This difficulty may be overcome by employing the very large symmetry of the system and mapping to other more tractable models. From our Monte Carlo simulation results and in the view that the effective repulsive interaction that leads to the breakup of the large clusters increases with larger $N$, the para-ferro transition is expected to occur only at $p=1$ when $N$ goes to infinity. The conjecture coincides with the rigorous proof that the ground state at $p=1$ is of saturated flat-band ferromagnetism (see Supplementary Material accompanying
this paper). In this limit $N\to\infty$, the ground state is paramagnetic as long as $p<1$ for both $1$D and $2$D cases.\\

6.\textbf{Conclusions}\\

We study the Hubbard models with SU$(N)$ spin symmetry and repulsive interaction on a type of decorated cubic lattices in one and two dimensions, where flat bands are present at the bottom of the band structures. When the filling density $p \equiv n / \mathcal{N} \le1$ with $n$ the number of particles and $\mathcal{N}$ the single particle flat band degeneracy, we prove that the ground states of the SU$(N)$ Hubbard model can be mapped to the geometrical configuration of an $N$-state Pauli-correlated percolation problem with a nontrivial weight. Based on this exact mapping, we solve for the ground states for 1D Tasaki lattice by a transfer-matrix method, and conclude that the system is paramagnetic at density  $p<1$ for arbitrary $N$. For 2D cases, we consider the SU$(N=3,4,10)$ symmetries as typical examples, and use the Metropolis Monte Carlo method to obtain the ground states numerically.
The para-ferro magnetic transition point $p_c$ is estimated to lay between $p_{-} < p_c < p_{+}$ with $p_-(N=3) = 0.64(1)$, $p_+(N=3) = 0.83(0)$, $p_-(N=4) = 0.65(1)$, $p_+(N=4) = 0.87(7)$, $p_-(N=10) = 0.69(4)$, $p_+(N=10) = 0.94(5)$. We also prove rigorously the existence of flat-band ferromagnetism for all values of $N \ge 2$ in arbitrary dimension, provided that the number of particles equals to the degeneracy of the flat band in the single-particle energy spectrum. The present problem reduces to the standard site percolation problem by taking $N=1$.\\

\textbf{Acknowledgements}\\

We would like to thank Zheng-Xin Liu, Deping Zhang and Zi-Xiang Li for helpful discussions. R.L. and W.Z. are supported by the National Key R\&D Program of China (Grant No. 2018YFA0306501), the National Natural Science Foundation of China (Grants No. 11434011, No. 11522436, and No. 11774425), the Beijing Natural Science Foundation (Grant No. Z180013), and the Research Funds of Renmin University of China (Grants No. 10XNL016 and No. 16XNLQ03). W.N. is supported in part by the National Key R\&D Program of China (Grant No. 2017YFB0405700), the National Natural Science Foundation of China (Grant No. 11704267) and start-up funding from Sichuan University (Grant No. 2018SCU12063).\\

\textbf{Author contributions}\\

W. Nie and W. Zhang conceived the project. All authors contributed equally to the analytical calculation. R. Liu and W. Nie performed the Metropolis Monte Carlo simulation for the 2D case. W. Nie and W. Zhang wrote the manuscript with inputs from all coauthors.\\

\appendix

In this Supplementary Material, we provide characterizations of the ground states for 1D and 2D Tasaki lattices and details of the transfer-matrix method applied to the 1D case.
\section{Characterization of the ground states}
\label{a}
In this section, we prove that the many-body ground states of the SU$(N)$ repulsive Hubbard model on lattices where the single-particle dispersion acquires flat-band ground states can be characterized by linear combinations of ferromagnetic clusters, provided the filling density $p \equiv n/\mathcal{N} \le 1$ with $n$ the number of particles and $\mathcal{N}$ the number of flat-band degeneracy. As the single-particle states in the non-dispersive flat band correspond to localized states, the value of $\mathcal{N}$ equals to the number of those localized states, which are also referred as trapping cells. We emphasize that although the Tasaki lattice are considered as a typical example, the proof is general for all flat-band lattices in arbitrary dimensions.

We denote the set of sites of the Tasaki lattices by $\Lambda$ and the undecorated hypercubic lattice by $V$, then $|V|=\mathcal{N}$. We rewrite the Hamiltonian in the main text as
\begin{eqnarray}
H = H_0+H_{\textrm{int}},
\label{H}
\end{eqnarray}
with $H_0=\sum_{i,j \in \Lambda}t_{i,j} \left( c_{i,\sigma}^{\dag}c_{j,\sigma}+\mbox{H.c.} \right)$, $H_{\textrm{int}}=U\sum\limits_{i} \sum\limits_{\sigma\neq\sigma'} n_{i,\sigma}n_{i,\sigma'}$.
When there exists a flat band, the eigenstates of single particle are localized states and can be represented as $\{a_{u,\sigma}^{\dagger}|0\rangle\}$, where $u=\{1,2,\cdots,\mathcal{N}\}$ denotes the trapping cells, $\sigma=\{1,2,\cdots,N\}$ labels the spin component, and $|0\rangle$ represents the vacuum state. Note that the trapping cells and the sites of the undecorated lattice have a one-to-one correspondence, such that $u$ also denotes the undecorated lattice sites. The expression of the creation operator $a_{u,\sigma}^{\dagger}$ reads
\begin{align}
a_{u,\sigma}^{\dagger}=\sum_{i\in \Lambda}\varphi_i^uc_{i,\sigma}^{\dagger},
\label{trapstate}
\end{align}
where $\varphi_i^u$ is defined as
\begin{equation}
\varphi_i^u=
\left\{
\begin{array}{rcl}
1, & & {i=u+\delta;}\\
-\sqrt{c}, & & {i=u;}\\
0, & & \textrm{otherwise.}
\end{array}
\right.
\end{equation}
Here, $\delta$ is the unit vector between the undecorated and decorated sites and $c$ is the coordination number of the corresponding undecorated lattice.
The commutation relation between $H_0$ and $a_{u,\sigma}^{\dagger}$ can be easily obtained
\begin{align}
\left[H_0, a_{u,\sigma}^{\dagger} \right] = \varepsilon_0a_{u,\sigma}^{\dagger}
\label{commute}
\end{align}
with $\varepsilon_0$ the single-particle eigenenergy of the flat band.

When the filling density $p <1$, the number of particles is less than that of the trapping cells, such that each trapping cell is occupied by at most one particle to avoid energy penalty of repulsion. Obviously, the ferromagnetic state
\begin{align}
|\Phi_{A,N}\rangle=\prod_{u\in A}a_{u,N}^{\dagger}|0\rangle
\label{ferrostate}
\end{align}
with all $n$ particles aligned at the spin $N$ component is a ground state of $H$ given by Eq.~(\ref{H}). Here, $A$ (with $|A|=n\leq\mathcal{N}$) is an arbitrary subset of the undecorated lattice $V$. It can be verified that $H_{0}|\Phi_{A,N}\rangle=n\varepsilon_0|\Phi_{A,N}\rangle$ and $H_{\textrm{int}}|\Phi_{A,N}\rangle=0$.

From the fully-polarized ferromagnetic state $|\Phi_{A,N}\rangle$, we can construct states composed of ferromagnetic clusters
 \begin{align}
|\Phi_{A,\{m_i\}}\rangle=\left\{ \prod_{i=1}^{M} \left[\prod\limits_{k=1}^{m_i}\mathbf{S}^{-}_{k,i}\right]' \right\} |\Phi_{A,N}\rangle,
\label{basis}
\end{align}
where $i = \{1, 2, \cdots M \}$ labels the linked clusters of occupied sites, $\mathbf{S}^{-}_{k,i}$ is one of the lowering operators of the complete lowering/raising operator set $\{\mathbf{S}^{\pm}_{1},\cdots,\mathbf{S}^{\pm}_{C_{N}^{2}}\}$ for the SU($N$) group acting on the \textit{i}-th cluster, $m_i$ is a non-negative integer among $\{0,1,\cdots,d_{\textrm{SU}(N)}(|C_i|)-1\}$ labeling the spin degeneracy, and $C_{N}^{2}$ is the binomial coefficient. The action of operators $[\mathbf{O}]'$ is arranged in such a way that $\mathbf{O}|\Phi_{A,N}\rangle\neq0$.
It can be verified that by selecting all possible combinations of $m_i$ and $\{\mathbf{S}^{\pm}_{k,i}\}$, we can obtain a number of $d_{\textrm{SU}(N)}(|C_i|)$ linearly independent degenerate states of the \textit{i}-th cluster.
Notice that these $d_{\textrm{SU}(N)}(|C_i|)$ degenerate ferromagnetic states can be represented by the Young tableau representation $:\bm{\sigma}_i:=D\sum_{P}\prod_{u_i\in C_i}a_{u_i,\sigma_{u_i}}^{\dagger}|0\rangle$, where $D$ is the normalized coefficient, $\bm{\sigma}_i=(\sigma_{u_1},\cdots,\sigma_{u_{|C_i|}})$ is the sector of spin configuration of the $i$-th cluster. The arrangement of spin components in $:\quad:$ is non-decreasing from left to right, and the summation over permutation $P$ takes account of all possible non-repeated spin configurations $(\sigma_{u_1},\cdots,\sigma_{u_{|C_i|}})$ of the \textit{i}-th cluster. Because of the commutation relation $[\mathbf{S}^{-}_{k,i},H]=0$, one can easily conclude that the states $|\Phi_{A,\{m_i\}}\rangle$ expressed in Eq.~(\ref{basis}) are also ground states of $H$. In the following, we will demonstrate that an arbitrary ground state of the SU$(N)$ Hubbard mode can be expanded by states defined by Eq.~(\ref{basis}) with various $A$, $\{m_i\}$ and $\{\mathbf{S}_{k,i}^{-}\}$.

An arbitrary ground state $|\Phi\rangle_{\textrm{GS}}$ of the SU$(N)$ Hubbard model can be represented with the aid of trapping cell states
\begin{widetext}
\begin{align}
|\Phi\rangle_{\textrm{GS}}=\sum_{A_1,\cdots A_N}f(A_1,\cdots A_N)\prod_{u_1\in A_1}a_{u_1,1}^{\dagger}\cdots\prod_{u_N\in A_N}a_{u_N,N}^{\dagger}|0\rangle,
\label{groundstate}
\end{align}
\end{widetext}
where $A_1,\cdots, A_N$ are subsets of $V$ satisfying the number constraints $|A_1|+|A_2|+\cdots+|A_N|=n$. For any given subsets $\{A_1,\cdots, A_N\}$, we can decompose them into sets of connected clusters $C_k$, such that $\bigcup\limits_{\sigma=1}^{N}A_\sigma=\bigcup\limits_{k=1}^{M}C_k$. By using the commutation relation $[ H_0, a_{u,\sigma}^{\dagger}] = \varepsilon_0a_{u,\sigma}^{\dagger}$, we obtain $H_0|\Phi\rangle_{\textrm{GS}}=n\varepsilon_0|\Phi\rangle_{\textrm{GS}}$.

The ground state $|\Phi\rangle_{\textrm{GS}}$ must satisfy another condition $H_{\textrm{int}}|\Phi\rangle_{\textrm{GS}}=0$, which is equivalent to $c_{i,\sigma}c_{i,\sigma'}|\Phi\rangle_{\textrm{GS}}=0$. The fermionic site operators $c_{i,\sigma}^\dagger$ and $c_{i,\sigma}$ can be expressed by the trapping cell operators $a_{u,\sigma}^{\dagger}$ and its corresponding adjoint. By introducing the Gramm matrix $G$
\begin{align}
(G)_{u,v}=\sum_{i\in\Lambda}\varphi_i^u\varphi_i^v,
\end{align}
and defining
\begin{align}
\kappa_{i}^{u}=\sum_{v\in V}(G^{-1})_{u,v}\varphi_i^v,
\end{align}
we have the completeness relation
\begin{eqnarray}
\sum\limits_{i\in\Lambda}\kappa_{i}^{u}\varphi_{i}^{v} &=& \delta_{u,v},\label{complete1}
\\
\sum\limits_{u\in V}\kappa_{i}^{u}\varphi_{j}^{u} &=& \delta_{i,j}-\psi_{ij},\label{complete2}
\end{eqnarray}
where
\begin{align}
\psi_{ij}=\psi_{ji}=\delta_{i,j}-\sum_{u,v\in V}\varphi_i^u(G^{-1})_{u,v}\varphi_j^v.
\end{align}
Thus, we can define the operators $b_{u,\sigma}$ adjoint to $a_{u,\sigma}^\dagger$
\begin{align}
b_{u,\sigma}=\sum\limits_{i\in\Lambda}\kappa_{i}^{u}c_{i,\sigma}
\end{align}
and the auxiliary operators
\begin{align}
d_{i,\sigma}=\sum\limits_{j\in\Lambda}\psi_{ij}c_{j,\sigma},
\end{align}
which satisfy the anti-commutation relations
\begin{eqnarray}
\left\{b_{u,\sigma},a^{\dagger}_{u',\sigma'}\right\} &=& \delta_{u,u'}\delta_{\sigma,\sigma'}\nonumber\\
\left\{d_{i,\sigma},a^{\dagger}_{u,\sigma'}\right\} &=& \left\{d^{\dagger}_{i,\sigma},b_{u,\sigma'}\right\}=0.
\end{eqnarray}
With the aid of the completeness relations (\ref{complete1}) and (\ref{complete2}), we obtain the expansion formulae
\begin{eqnarray}
c^{\dagger}_{i,\sigma} &=& \sum\limits_{u\in V}\kappa_{i}^{u}a^{\dagger}_{u,\sigma}+d^{\dagger}_{i,\sigma},\label{expansion1}\\
c_{i,\sigma} &=& \sum\limits_{u\in V}\varphi_{i}^{u}b_{u,\sigma}+d_{i,\sigma}
\label{expansion2}
\end{eqnarray}
for arbitrary $i\in\Lambda$ and $\sigma=1,2, \cdots,N$. By substituting Eq.~(\ref{expansion2}) into the ground state condition $c_{i,\sigma}c_{i,\sigma'}|\Phi\rangle_{\textrm{GS}}=0$, and employ the quasi-locality and local connectivity conditions~\cite{1,2,3,4} satisfied by the Tasaki lattices, we can finally obtain two constraints for the coefficients $f(A_1,\cdots A_N)$:

(i) One must have
\begin{align}
f(A_1,\cdots, A_N)=0
\label{require1}
\end{align}
if $A_\sigma \cap A_{\sigma^\prime} \neq\emptyset$ for any $\sigma \neq  \sigma^\prime$. This indicates that in each particle distribution $\{A_1,\cdots, A_N\}$ in the ground states, the subsets $A_\sigma$ and $A_{\sigma^\prime}$ do not overlap. In other words, a lattice site is occupied by at most one particle.

(ii) For any two subsets $\{A_1,\cdots, A_N\}$ and $\{A_1^\prime,\cdots, A_N^\prime\}$ which can be decomposed into the same set of clusters $\bigcup\limits_{\sigma=1}^{N}A_\sigma=\bigcup\limits_{\sigma^\prime = 1}^{N}A_{\sigma^\prime}^\prime=\bigcup\limits_{k=1}^{M}C_k$ and satisfy $A_\sigma \cap A_{\sigma^\prime} =A_{\sigma}^\prime \cap A_{\sigma^\prime}^\prime = \emptyset$ for any $\sigma \neq \sigma^\prime$, if the number of particles of the $\sigma$-th spin component are the same within each cluster, i.e.,
\begin{align}
\left\vert A_\sigma \cap C_k \right\vert = \left\vert A_{\sigma}^\prime \cap C_k \right\vert
\label{condition2}
\end{align}
for all $\sigma$ and $k$, the coefficients associated with the two subsets are equivalent
\begin{align}
f(A_1,\cdots, A_N)= f(A_1^\prime,\cdots, A_N^\prime).
\label{require2}
\end{align}
Thus, the coefficients $f(A_1,\cdots, A_N)$ are symmetric under permutation of particles within an individual cluster. The summation of all possible permutation states then resembles the ferromagnetic state obtained from the fully-polarized cluster after a sequence of spin lowering operator.

The constraints (i) and (ii) prove that any ground state $|\Phi\rangle_{\textrm{GS}}$ is a linear combination of the states given by Eq.~(\ref{basis}). This characterization of ground states is important when mapping to a percolation problem as well as using the transfer-matrix method.

Before concluding this section, we emphasize that for the special case with $n = \mathcal{N}$, the flat-band is $1/N$ filled and there can only exist one cluster with size equals to the number of trapping cells. Thus, the states by linear combination of Eq.~(\ref{basis}) are fully ferromagnetic states with a spin degeneracy associated with the SU$(N)$ symmetry, which proves the flat-band ferromagnetism of the ground state (When we are submitting this paper, we are informed that flat band ferromagnetism of SU($N$) Hubbard model at filling $p=1$ was mentioned in the talk of H. Katsura: \url{https://pcs.ibs.re.kr/PCS_Workshops/PCS_Flatband_Networks_in_Condensed_Mater_and_Photonics.html}).
Notice that this statement is valid for arbitrary $N$ in any dimensions.

\section{Transfer-matrix method for 1D Tasaki lattice}
\label{b}
As discussed in the main text, the populating rules of $N$-component particles over $\mathcal{N}$ traps can be expressed by a transfer matrix $\mathbf{T}$, which reads
\begin{align}
\mathbf{T}&=
\left(
  \begin{array}{ccccc}
    T(0,0) & T(0,1) & T(0,2) & \cdots & T(0,N)\\
    T(1,0) & T(1,1) & T(1,2) & \cdots & T(1,N)\\
    T(2,0) & T(2,1) & T(2,2) & \cdots & T(2,N)\\
    \vdots & \vdots & \vdots & \ddots & \vdots\\
    T(N,0) & T(N,1) & T(N,2) & \cdots & T(N,N)\\
  \end{array}
\right)\nonumber\\&=
\left(
  \begin{array}{ccccc}
    1 & 1 & 1 & \cdots & 1\\
    z & z & z & \cdots & z\\
    z & 0 & z & \cdots & z\\
    \vdots & \vdots & \vdots & \ddots & \vdots\\
    z & 0 & 0 & \cdots & z\\
  \end{array}
\right)
\end{align}
with $z=e^{\mu}$.
The matrix element $T(\sigma_i,\sigma_{i+1})$ takes the value $1$ ($z$) if the site $i$ is empty (occupied) for spin configurations satisfying the two rules defined in the main text.
The contribution of all possible configurations to the
grand-canonical partition function can be calculated as
\begin{align}
\Xi(z,\mathcal{N})=\textrm{Tr} \,\mathbf{T}^\mathcal{N}.
\end{align}
To calculate some physical quantities such as the average occupation number of a given site $\langle n_i\rangle$, we introduce the $\mathbf{N}$-matrix,
\begin{eqnarray}
\mathbf{N}=
\left(
  \begin{array}{ccccc}
    0 & 0 & 0 & \cdots & 0\\
    0 & 1 & 0 & \cdots & 0\\
    0 & 0 & 1 & \cdots & 0\\
    \vdots & \vdots & \vdots & \ddots & \vdots\\
    0 & 0 & 0 & \cdots & 1\\
  \end{array}
\right),
\end{eqnarray}
and we have
\begin{eqnarray}
\langle n_i\rangle=\frac{\textrm{Tr} \, \mathbf{T}^\mathcal{N}\mathbf{N}}{\textrm{Tr}\,\mathbf{T}^\mathcal{N}}=p=\frac{n}{\mathcal{N}}.
\end{eqnarray}
\begin{figure*}
\centering
\includegraphics[width=6.5in]{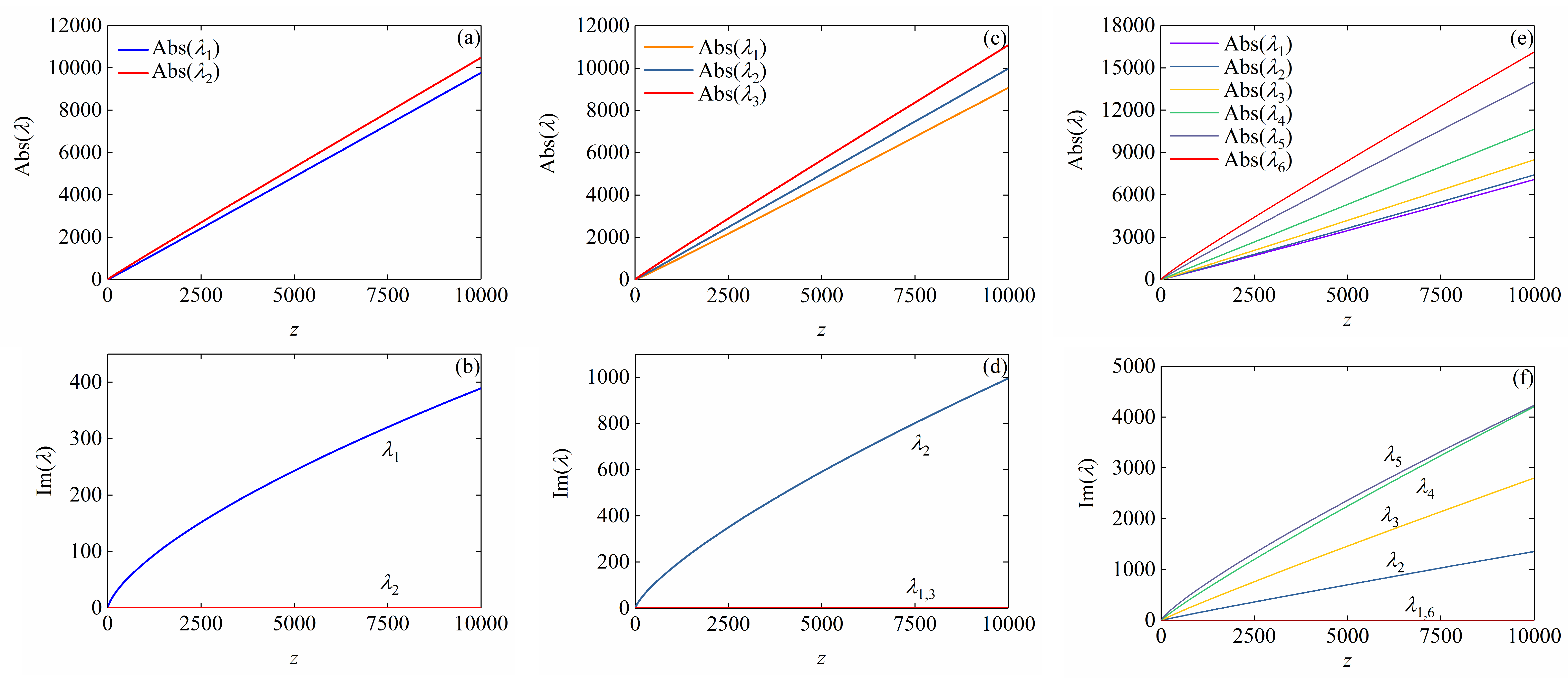}
\caption{(Color online) The modulus (top row)and imaginary part (bottom row) of roots $\lambda_i$ of Eq.~(\ref{equation}) for the cases of (a-b) $N=3$, (c-d) $N=4$, and (e-f) $N=10$. Here we show only the roots with positive imaginary part as the ones with negative imaginary part are simply the corresponding complex conjugates. Note that the root with maximal modulus (red solid) is real in all cases.}
\label{rootdis}
\end{figure*}

Besides, we introduce the $\mathbf{S},\mathbf{C},\mathbf{F}$-matrix,
\begin{align}
\mathbf{S}&=
\left(
  \begin{array}{ccccc}
    0 & 1 & 1 & \cdots & 1\\
    0 & 0 & 0 & \cdots & 0\\
    0 & 0 & 0 & \cdots & 0\\
    \vdots & \vdots & \vdots & \ddots & \vdots\\
    0 & 0 & 0 & \cdots & 0\\
  \end{array}
\right),\quad
\mathbf{C}=
\left(
  \begin{array}{ccccc}
    0 & 0 & 0 & \cdots & 0\\
    0 & z & z & \cdots & z\\
    0 & 0 & z & \cdots & z\\
    \vdots & \vdots & \vdots & \ddots & \vdots\\
    0 & 0 & 0 & \cdots & z\\
  \end{array}
\right),\nonumber\\
\mathbf{F}&=
\left(
  \begin{array}{ccccc}
    0 & 0 & 0 & \cdots & 0\\
    z & 0 & 0 & \cdots & 0\\
    z & 0 & 0 & \cdots & 0\\
    \vdots & \vdots & \vdots & \ddots & \vdots\\
    z & 0 & 0 & \cdots & 0\\
  \end{array}
\right)
\end{align}
which means a sequence starting with an empty site, both neighboring traps in a sequence being occupied, and the last site of a sequence being followed by an empty one respectively.
Note that the $\mathbf{C}$-matrix satisfies
\begin{eqnarray}
\mathbf{C}^n=z^n
\left(
  \begin{array}{ccccc}
    0 & 0 & 0 & \cdots & 0\\
    0 & 1 & n & \cdots & \frac{n(n+1)(n+2)\cdots(n+N-2)}{(N-1)!}\\
    0 & 0 & 1 & \cdots & \frac{n(n+1)(n+2)\cdots(n+N-3)}{(N-2)!}\\
    \vdots & \vdots & \vdots & \ddots & \vdots\\
    0 & 0 & 0 & \cdots & 1\\
  \end{array}
\right).
\end{eqnarray}

For convenience, we diagonalize the $\mathbf{T}$-matrix with $\mathbf{\Lambda}=\mathbf{X}^{-1}\mathbf{T}\mathbf{X}$, where the diagonal matrix reads
\begin{align}
\mathbf{\Lambda}=
\left(
  \begin{array}{ccccc}
    0 & 0 & 0 & \cdots & 0\\
    0 & \lambda_1 & 0 & \cdots & 0\\
    0 & 0 & \lambda_2 & \cdots & 0\\
    \vdots & \vdots & \vdots & \ddots & \vdots\\
    0 & 0 & 0 & \cdots & \lambda_{N}\\
  \end{array}
\right)
\end{align}
with $\lambda_1,\lambda_2,\cdots,\lambda_N$ the non-zero and non-repeated roots of the following equation
\begin{align}
(\lambda-z)^N-\lambda^{N-1}=0,\label{equation}
\end{align}
and the transformation matrix
\begin{align}
\mathbf{X}=
\left(
  \begin{array}{ccccc}
    -1 & \frac{\lambda_1-z}{z} & \frac{\lambda_2-z}{z} & \cdots & \frac{\lambda_N-z}{z}\\
    0 & \lambda_1-z & \lambda_2-z & \cdots & \lambda_N-z\\
    0 & \frac{(\lambda_1-z)^2}{\lambda_1} & \frac{(\lambda_2-z)^2}{\lambda_2} & \cdots & \frac{(\lambda_N-z)^2}{\lambda_N}\\
    \vdots & \vdots & \vdots & \ddots & \vdots\\
    1 & 1 & 1 & \cdots & 1\\
  \end{array}
\right).
\end{align}

Before proceeding the calculation, it is necessary to analyze the properties of these $N$ roots. It can be verified that if $\lambda_i$ is a root, then $\lambda_{i}^{*}$ is also a root. In other words, there exits a pair of complex conjugate roots. Besides, we point out that the root with the maximal modulus is always real as demonstrated in Fig.~\ref{rootdis}.

The cluster distribution $n(l)$ can be obtained as
\begin{align}
\lim\limits_{\mathcal{N}\to\infty}n(l,\text{SU}(N))&=\lim\limits_{\mathcal{N}\to\infty}\frac{\textrm{Tr}\,\mathbf{T}^{\mathcal{N}-l-1}\mathbf{S}\mathbf{C}^{l-1}\mathbf{F}}{\textrm{Tr}\,\mathbf{T}^\mathcal{N}}\nonumber\\
&\hspace{-1.5cm}=\lim\limits_{\mathcal{N}\to\infty}\frac{\textrm{Tr}\,\mathbf{X}\mathbf{\Lambda}^{\mathcal{N}-l-1}\mathbf{X^{-1}}\mathbf{S}\mathbf{C}^{l-1}\mathbf{F}}{\textrm{Tr}\,\mathbf{T}^\mathcal{N}}.
\end{align}
We can find that the elements of $\mathbf{S}\mathbf{C}^{l-1}\mathbf{F}$ are all zeros except
$$(\mathbf{S}\mathbf{C}^{l-1}\mathbf{F})_{1,1}=z^l\sum\limits_{i=2}^{n+1}\sum\limits_{j=2}^{n+1}\left(\frac{\mathbf{C}^{l-1}}{z^{l-1}}\right)_{i,j}.$$ Thus, in the thermodynamic limit $\mathcal{N}\to\infty$ we obtain
\begin{widetext}
\begin{align}
\lim\limits_{\mathcal{N}\to\infty}n(l,\text{SU}(N))&=\frac{(\mathbf{X}\mathbf{\Lambda}^{\mathcal{N}-l-1}\mathbf{X}^{-1})_{1,1}(\mathbf{S}\mathbf{C}^{l-1}\mathbf{F})_{1,1}}{\lambda_{\text{max}}^{\mathcal{N}}}\nonumber\\
&=\lim\limits_{\mathcal{N}\to\infty}\frac{(\mathbf{X})_{1,\text{max}}\lambda_{\text{max}}^{\mathcal{N}-l-1}(\mathbf{X}^{-1})_{\text{max},1}z^l}{\lambda_{\text{max}}^{\mathcal{N}}}
\sum\limits_{i=2}^{N+1}\sum\limits_{j=2}^{N+1}\left(\frac{\mathbf{C}^{l-1}}{z^{l-1}}\right)_{i,j}
\nonumber\\
&=f_{\text{SU}(N)}(z)\frac{(l+N-1)!}{(N-1)!  l!}\left(\frac{z}{\lambda_{\text{max}}(z)}\right)^l, \label{acluster}
\end{align}
\end{widetext}
where $f_{\text{SU}(N)}(z)=\frac{(\mathbf{X})_{1,\text{max}}(\mathbf{X}^{-1})_{\text{max},1}}{\lambda_{\text{max}}(z)}$,
$\lambda_{\text{max}}(z)$ is the root with the maximal modulus of the $N$ roots, and $\text{max}(\neq1)$ is the index which satisfies $(\mathbf{\Lambda})_{\text{max},\text{max}}=\lambda_{\text{max}}$.
We emphasize again that $\lambda_{\text{max}}$ is real and if not, we will obtain a cluster distribution related to $\cos[\mathcal{N}\text{arg}(\lambda_{\cdots})]$, which will oscillate in the limit of $\mathcal{N}\to\infty$ rather than converging to a fixed value.

The percolation transition occurs when the condition ${z_c/\lambda_{\text{max}}(z_c)}=1$ is satisfied. To determine the transition point $z_c$, we rewrite the roots of Eq.~(\ref{equation}) as $\lambda=\alpha(z)z$ with complex $\alpha(z)$, such that the percolation transition takes place when $\alpha = 1$. By substituting the representation of $\lambda$ back into Eq.~(\ref{equation}), we obtain
\begin{align}
z=\left(\frac{\alpha_i}{\alpha_i-1}\right)^{N-1}\frac{1}{\alpha_i-1}
\label{zalpha}
\end{align}
with $i=1,2,\cdots,N$. From the expression above, one can easily find that $z \to + \infty$ when $\alpha \to 1$. Indeed, it can be verified that there exists at least one real $\alpha(z)$ satisfying Eqs.~(\ref{equation}) and (\ref{zalpha}) for arbitrary $z>0$, and the absolute value of $\alpha$ is no less than 1 (only if $z \to +\infty$). Hence the percolation transition occurs at $z_c \to +\infty$. We further point out that at the critical point $z_c \to +\infty$, the real component of $\alpha_i$ approaches unity while the imaginary part goes to zero.

In order to determine the results in terms of the average occupation number $p$, in the thermodynamic limit we have,
\begin{align}
\lim\limits_{\mathcal{N}\to\infty}p(z)&=\lim\limits_{\mathcal{N}\to\infty}\frac{\textrm{Tr}\,\mathbf{T}^\mathcal{N}\mathbf{N}}{\textrm{Tr}\,\mathbf{T}^\mathcal{N}}=\lim\limits_{\mathcal{N}\to\infty}\frac{\textrm{Tr}\,\mathbf{X}\mathbf{\Lambda}^\mathcal{N}\mathbf{X}^{-1}\mathbf{N}}{\textrm{Tr}\,\mathbf{X}\mathbf{\Lambda}^\mathcal{N}\mathbf{X}^{-1}}\nonumber\\
&=\lim\limits_{\mathcal{N}\to\infty}\frac{\textrm{Tr}\,\mathbf{X}\mathbf{\Lambda}^\mathcal{N}\mathbf{X}^{-1}-(\mathbf{X}\mathbf{\Lambda}^\mathcal{N}\mathbf{X}^{-1})_{11}}{\textrm{Tr}\,\mathbf{X}\mathbf{\Lambda}^\mathcal{N}\mathbf{X}^{-1}}\nonumber\\
&=1-\lim\limits_{\mathcal{N}\to\infty}\frac{(\mathbf{X}\mathbf{\Lambda}^\mathcal{N}\mathbf{X}^{-1})_{11}}{\textrm{Tr}\,\mathbf{X}\mathbf{\Lambda}^\mathcal{N}\mathbf{X}^{-1}}\nonumber\\
&=1-(\mathbf{X})_{1,\text{max}}(\mathbf{X}^{-1})_{\text{max},1}.
\label{pz}
\end{align}
We further let $z\to+\infty$ and obtain with the aid of Eq.~(\ref{zalpha}) that $\lim\limits_{z\to\infty}(\lambda_i-z)/{z}=\lim\limits_{z\to\infty}(\alpha_i(z)-1)\to0$ for $i=1,2,\cdots,N$. Thus we have $(\mathbf{X})_{1,i}=-1\delta_{i,1}$ and $(\mathbf{X}^{-1})_{i,1}=-1\delta_{i,1}$ for $i=1,2,\cdots,N+1$, which leads to $(\mathbf{X}\mathbf{\Lambda}^\mathcal{N}\mathbf{X}^{-1})_{11}/(\textrm{Tr}\,\mathbf{X}\mathbf{\Lambda}^\mathcal{N}\mathbf{X}^{-1})\to0$ and $p\to 1$.

\section{Supplementary material for 2D Monte-Carlo simulation}

\begin{figure}
	\begin{center}
		\includegraphics[width=8cm]{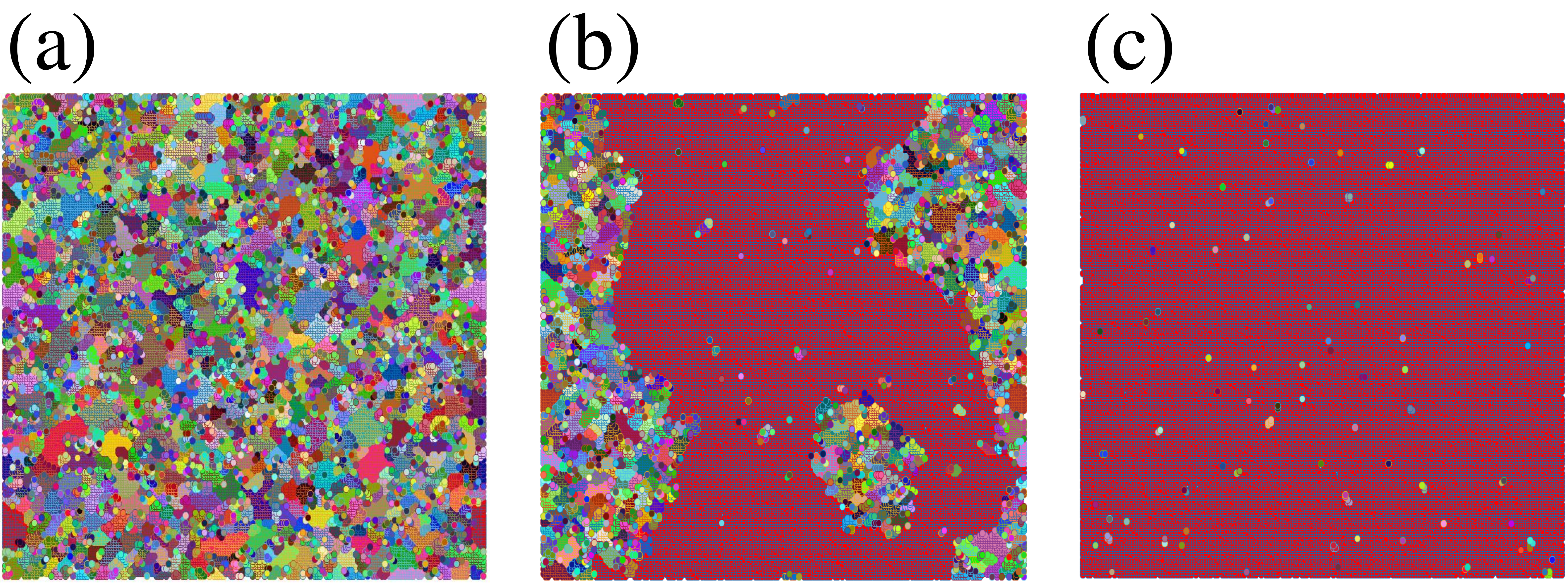}
	\end{center}
	\caption{\label{snapshot}(Color online) Snapshots of configurations for Pauli-correlated percolation corresponding to the SU($3$) Hubbard model, at concentrations of (a) $p_1=0.64$ (paramagnetic), (b) $p_2=0.74$ (phase-separated), and (c) $p_3=0.83$ (ferromagnetic). The sample size is $200\times200$}.
\end{figure}

To show the features of the ground state at different particle concentrations in canonical simulations, we plot as a typical example the snapshots of Pauli-correlated percolation configurations for SU($3$) Hubbard model in Fig.~\ref{snapshot}. When particle density $p_1=0.64<p_c$ as shown in Fig.~\ref{snapshot}a, the ground state is paramagnetic with no macroscopic clusters. When particle density exceeds $p_c$ as illustrated in Fig.~\ref{snapshot}b, the ground state becomes ferromagnetic with macroscopic domains which are phase-separated from each other. If the filling factor is further increased, ferromagnetic phase spans the entire system as depicted in Fig.~\ref{snapshot}c.

\section{Connection to the standard percolation problem in the limit of $N=1$}
From Eq.(\ref{acluster}) of the Supplementary Material, we obtain the cluster distribution for 1D Tasaki lattice in the limit of $N=1$,
\begin{align}
\lim\limits_{\mathcal{N}\to\infty}n(l,\text{SU}(1))=\frac{z^{l}}{(1+z)^{l+2}}.
\label{aclusterN1}
\end{align}
With the aid of Eq. (\ref{pz}) of the Supplementary Material, we can further get the average occupation number $p(z)$ in the thermodynamic limit,
\begin{align}
\lim\limits_{\mathcal{N}\to\infty}p(z)=1-\frac{1}{1+z}.
\end{align}
The result above determines the fugracity $z=\frac{p}{1-p}$. By substituting this expression
into Eq. (\ref{aclusterN1}) in the Supplementary Material, we finally obtain
\begin{align}
\lim\limits_{\mathcal{N}\to\infty}n(l,\text{SU}(1))=p^l(1-p)^2,
\end{align}
which is consistent with the standard percolation~\cite{5}.

For 2D case, according to Eq.~(2) in the main text, the spin degeneracy $d_{\text{SU}(N)}=1$ in the limit of $N=1$. Thus, we recover the standard percolation problem.

\end{document}